\documentclass[a4paper,11pt]{article}

\usepackage{jheppub}

\usepackage{subfigure}
\usepackage{epsfig}
\usepackage{amssymb,amsmath}

\title{A perturbative QCD study of dijets in p+Pb collisions at the LHC}

\author[a]{Kari J. Eskola,}
\author[a]{Hannu Paukkunen,}
\author[b]{Carlos A. Salgado}

\affiliation[a]{Department of Physics, University of Jyv\"askyl\"a, P.O. Box 35, FI-40014, Finland, and \\
Helsinki Institute of Physics, University of Helsinki, P.O. Box 64, FI-00014, Finland}
\affiliation[b]{Departamento de F\'\i sica de Part\'\i culas and IGFAE, Universidade de Santiago de 
Compostela, E-15782 Santiago de Compostela, Galicia, Spain}

\emailAdd{kari.eskola@jyu.fi}
\emailAdd{hannu.paukkunen@jyu.fi}
\emailAdd{carlos.salgado@usc.es}

\abstract{Inspired by the recent measurements of the CMS collaboration,  
we report a QCD study of dijet production in proton+lead collisions at the LHC 
involving large-transverse-momentum jets, $p_T \gtrsim 100$ GeV. Examining the
inherent uncertainties of the next-to-leading order perturbative QCD calculations
and their sensitivity to the free proton parton distributions (PDFs), we observe a rather small,
typically much less than 5\% clearance for the shape of the dijet rapidity distribution
within approximately 1.5 units around the midrapidity. Even a more stable observable
is the ratio between the yields in the positive and negative dijet rapidity, for which 
the baseline uncertainty can be made negligible by imposing a symmetric jet rapidity 
acceptance. Both observables prove sensitive to the nuclear modifications of the gluon 
distributions, the corresponding uncertainties clearly exceeding the estimated baseline
uncertainties from the free-proton PDFs and scale dependence. 
From a theoretical point of view, these observables are therefore very suitable for 
testing the validity of the collinear factorization and have a high potential to 
provide precision constraints for the nuclear PDFs.}

\begin{document}

\maketitle
\flushbottom

\section{Introduction}

The recently completed proton+lead run at the LHC with $\sqrt{s} = 5.02 \, {\rm TeV}$ proton-nucleon center-of-mass energy
\cite{Loizides:2013nka,Salgado:2011wc} has opened a possibility to study various large-transverse-momentum ($p_T$) processes at an unprecedented
energy realm in nuclear collider experiments. Along with the large center-of-mass energy, processes like 
production of on-shell heavy gauge bosons that have routinely been measured e.g. in proton+antiproton collisions at the
Tevatron and in proton+proton collisions at the LHC, have now become measurable also in proton+nucleus collisions.

Collinear factorization \cite{Collins:1989gx,Collins:1985ue} constitutes a coherent theoretical framework providing
a baseline against which the LHC data from various high-$p_T$ processes in proton+lead collisions can be contrasted
\cite{Albacete:2013ei}. The non-perturbative input needed in such calculations are the parton distribution functions (PDFs)
of the free proton \cite{Martin:2009iq,Lai:2010vv,Alekhin:2012ig,Forte:2013wc,Ball:2012cx} and those of the bound nucleons,
the nuclear PDFs. These have been obtained in global analyses, which in the nuclear case \cite{Eskola:2009uj,deFlorian:2011fp,Hirai:2007sx,Schienbein:2009kk,Kovarik:2013sya} use, in different combinations, neutral-current lepton+nucleus deeply inelastic scattering (DIS) data, Drell-Yan dilepton data in proton+nucleus collisions, inclusive pion production in deuterium+gold collisions, and charged-current neutrino+nucleus DIS data --- for a recent review, see~\cite{Eskola:2012rg}.
For an ongoing discussion concerning the neutrino DIS, see Refs.~\cite{Kovarik:2010uv,Paukkunen:2010hb,Paukkunen:2013grz}.
The extent to which the nuclear PDFs have been tested  is, however, both kinematically and process-wize still much more
limited than in the case of the free proton PDFs. Further tests of the universality of the nuclear PDFs, and novel 
constraints for them as well, are expected from the forthcoming LHC nuclear data.

In fact, there are already some evidence supporting the factorization at the LHC nuclear collisions: even
in lead+lead collisions the measurements for large-$p_T$ leptons \cite{Chatrchyan:2012nt}, high-mass dilepton-pairs \cite{Chatrchyan:2011ua,Aad:2012ew}
and high-$p_T$ photons \cite{Chatrchyan:2012vq,Wilde:2012wc} are 
consistent with the pQCD expectations 
\cite{Paukkunen:2010qg,Chatterjee:2013naa,Klasen:2013mga}, although the experimental uncertainties are still too large to make
decisive conclusions. The $p_T$
dependence of the nuclear modification of the charged hadron production from the p+Pb pilot run measured by the
ALICE collaboration \cite{ALICE:2012mj} is as well compatible with the nuclear PDFs \cite{Helenius:2012wd}. 
However, deviations from the factorization-based calculations in proton+lead collisions are predicted to follow, especially
in the low-$p_T$ region, from e.g. gluon saturation, parton energy loss and final state
interactions \cite{Albacete:2013tpa,Albacete:2013ei,Armesto:2013fca}. Also the search for such phenomena makes 
the LHC proton+lead data particularly interesting \cite{Loizides:2013nka}.  

Here, inspired by the preliminary results reported by the CMS collaboration \cite{CMSprel},
we consider inclusive dijet production in proton+lead collisions. This is not
the first time that jet-like particle bursts have been observed in proton+nucleus collisions:
Earlier such measurements come from fixed-target experiments in Fermilab \cite{Stewart:1990wa,Alverson:1993yc}
and HERA \cite{Golubkov:2006ty} with a fairly small center-of-mass energy $\sqrt{s} \backsimeq 40 \, {\rm GeV}$ 
and consequently involving rather small-$p_T$ jets, $p_T < 15 \, {\rm GeV}$.
More recently, the PHENIX \cite{Sahlmueller:2012ru} and STAR \cite{Kapitan:2010fb} collaborations at
the BNL-RHIC have successfully reconstructed jets in $\sqrt{s} = 200 \, {\rm GeV}$
d+Au collisions up to $p_T = 45 \, {\rm GeV}$. In what follows, we will discuss the size of the 
theoretical uncertainty of the pQCD calculations by comparing the
leading order (LO) calculations with next-to-leading order (NLO) ones, their sensitivity to the scale
variations, and to the PDF uncertainties. These uncertainties are compared to the expected magnitude 
of the nuclear effects from different global fits of nuclear PDFs in order to see whether they could be resolved.
For earlier dijet studies in this direction, see e.g.~\cite{He:2011sg}.

\section{The Framework}

\subsection{Definition of the jet cross section}

In this paper we consider the process
$
{\rm lead} + {\rm proton} \rightarrow {\rm dijet} + X,
$
at $\sqrt{s} = 5.02 \, {\rm TeV}$ proton-nucleon center-of-mass energy using the framework of 
collinear factorization. 
We will perform all the perturbative QCD calculations at the NLO level but, for simplicity,
let us first illustrate the situation using the LO formalism.
In this approximation, a dijet event consists of two partons with
(pseudo)rapidities $\eta_1$ and $\eta_2$, carrying equal transverse momentum $p_T$,
and the corresponding cross section can be written as \cite{Owens:1986mp}
\begin{equation}
\frac{d\sigma_{\rm dijet}}{dp_T^2 d\eta_1 d\eta_2} = \frac{1}{16\pi s^2} \sum_{ijkl}
\frac{f_i^{\rm Pb}(x_1,Q^2)}{x_1} \frac{f_j^{p}(x_2,Q^2)}{x_2}
\left| \mathcal{M}_{ij\rightarrow kl}\right|^2,
\end{equation}
where $f_i^{\rm Pb}(x_1,Q^2)$ and $f_i^{\rm p}(x_2,Q^2)$ are the PDFs, and $\left| \mathcal{M}_{ij\rightarrow kl}\right|^2$ 
is the squared matrix element for the partonic process $ij \rightarrow kl$. The momentum fractions $x_1$ and $x_2$ are given by
\begin{equation}
x_1 = \frac{p_T}{\sqrt s} \left( e^{\eta_1}+e^{\eta_2}\right), \quad x_2 = \frac{p_T}{\sqrt s} \left( e^{-\eta_1}+e^{-\eta_2}\right).
\end{equation}
The experimental
dijet data are often presented in bins of dijet invariant mass $M_{\rm dijet}$, given in LO by 
\begin{equation}
M_{\rm dijet}^2 = 2p_T^2(1+\cosh(\eta_1-\eta_2)), 
\end{equation}
 and, especially in symmetric proton+proton collisions, in a rapidity variable like $|y|_{\rm max}=\max(|y_1|,|y_2|)$ \cite{Chatrchyan:2012bja}
or $|y_1 - y_2|/2$ \cite{Aad:2011fc}. However, in order to plainly probe the $x$ dependence of the nuclear modifications
of the PDFs, a better variable is 
\begin{equation}
\eta_{\rm dijet} \equiv (\eta_1 + \eta_2)/2,  
\end{equation}
which we refer to as the dijet ''pseudorapidity''.\footnote{
Note that in LO, $\eta_{\rm dijet}$ above is strictly speaking the dijet \textit{rapidity} since for massless jets their
rapidities and pseudorapidities coincide. Beyond LO, $\eta_{\rm dijet}$ is neither the rapidity nor pseudorapidity.
We make this choice of variable in order to match the one in \cite{CMSprel}. However, our conclusion will
generally apply also for the dijet rapidity $y_{\rm dijet} \equiv (y_1 + y_2)/2$ distributions.
}
This is because, in the leading order, the momentum fractions for a given invariant mass $M_{\rm dijet}$ become simply
\begin{equation}
x_1 = \frac{M_{\rm dijet}}{\sqrt{s}} e^{\eta_{\rm dijet}}, \quad x_2 = \frac{M_{\rm dijet}}{\sqrt{s}} e^{-\eta_{\rm dijet}},
\label{eq:mfrac}
\end{equation}
and the data binned in $M_{\rm dijet}$ and $\eta_{\rm dijet}$ provide therefore a cleaner way to learn about the $x$
dependence of the nuclear PDFs. In practice, we will compute the dijet distributions in bins $\Delta \eta_{\rm dijet}$
of the dijet pseudorapidity, imposing a lower $p_T$ cut, $p_T^{\rm min}$, and
a pseudorapidity acceptance, $\Delta \eta$, for the individual jets, so that
\begin{eqnarray}
\frac{d\sigma_{\rm dijet}}{d\eta_{\rm dijet}}  &=& \frac{1}{\Delta \eta_{\rm dijet}} \int dp_T^2 d\eta_1 d\eta_2 \left( \frac{d\sigma_{\rm dijet}}{dp_T^2 d\eta_1 d\eta_2} \right)
\theta(p_T\ge p_T^{\rm min})\theta(\eta_1\in\Delta \eta)\theta(\eta_2\in\Delta \eta) \nonumber\\
&&\theta(|\eta_{\rm dijet}-\frac{\eta_1+\eta_2}{2}|\le\frac{\Delta\eta_{\rm dijet}}{2}).
\end{eqnarray}
 A finer binning in $M_{\rm dijet}$ will not bring much
additional insight to our main findings, as we will see.

We perform the calculations through NLO using a jet code \texttt{MEKS} \cite{Gao:2012he} which we have 
adapted for nuclear collisions and boosted to the laboratory frame. This program has its roots in the original
Ellis-Kunszt-Soper (EKS) routine \cite{Kunszt:1992tn,Ellis:1992en} for
inclusive jets and dijets. In order to facilitate the comparison with the forthcoming experimental measurements
we choose to present our calculations in the laboratory (collider) frame and use the same kinematical cuts
as the CMS experiment \cite{CMSprel} --- considering these as typical for the LHC.
The lead+proton run was accomplished by 
colliding lead ions of  $E_{\rm lead} = (82/208) \times 4 \,{\rm TeV} \approx 1.58 \, {\rm TeV}$ per nucleon energy onto
a beam of $E_{\rm p} = 4 \, {\rm TeV}$ protons.
Due to the unequal
energies of the colliding nucleons the center-of-mass midrapidity shifts by
\begin{equation}
\eta_{\rm shift} \equiv 0.5 \log\left( E_{\rm Pb}/E_{\rm p} \right) \approx -0.465.
\end{equation}
The individual jets are required to stay within the pseudorapidity acceptance of six units, $|\eta^{\rm leading, subleading}| < 3$,
and the transverse momenta carried by the leading and subleading jet are restricted by the conditions $p_T^{\rm leading} \ge p_T^{\rm min, leading} = 120 \, {\rm GeV}$
and $p_T^{\rm subleading} \ge p_T^{\rm min, subleading} = 30 \, {\rm GeV}$. In addition, the jets are required to be clearly
 separated in the relative azimuthal angle, $\Delta \phi > 2\pi/3$. The partons are assembled to jets according to the
anti-$k_T$ algorithm \cite{Cacciari:2008gp} with the distance parameter $R=0.3$, using the 4-vector recombination scheme.
Our calculation does not include any non-perturbative corrections due to e.g. effects of hadronization
or underlying event \cite{Dooling:2012uw,Dasgupta:2007wa}, which are sometimes estimated by using Monte-Carlo
event generators \cite{Corcella:2000bw,Sjostrand:2006za,Gleisberg:2008ta}.
However, with the requirement of a large leading-jet $p_T$, such non-perturbative effects are expected
to be suppressed \cite{Aad:2011fc,Dooling:2012uw}. In addition, for the
observables which we find especially suitable for testing the factorization (ratios of cross sections
with the same $R$) we would expect the non-perturbative corrections to be even more reduced.

\subsection{Nuclear Modifications in PDFs}

The PDFs $f_i^A$ for a nucleus with a mass number $A$ which we use in our calculations are linear
combinations of bound proton $f_i^{{ p},A}$ and bound neutron $f_i^{{n},A}$ PDFs,
\begin{equation}
f_i^A(x,Q^2) = \left( \frac{Z}{A} \right) f_i^{p,A}(x,Q^2) + \left( \frac{N}{A} \right) f_i^{n,A}(x,Q^2),
\end{equation}
where $Z$ and $N$ are the number of protons and neutrons correspondingly (here $Z=82$ and $N=126$). The PDFs of a bound
proton are obtained from the free proton PDFs $f_i^p$ and the corresponding nuclear modification 
factor $R_i^A(x,Q^2)$ as
\begin{equation}
f_i^{p,A}(x,Q^2) = R_i^A(x,Q^2) f_i^{p}(x,Q^2).
\label{eq:nmod}
\end{equation}
The bound neutron PDFs are obtained --- neglecting here all the QED
effects \cite{Martin:2004dh,Ball:2013hta} --- by an interchange of up and down flavors, $f_u^{n,A}(x,Q^2) = f_d^{p,A}(x,Q^2)$
and $f_d^{n,A}(x,Q^2) = f_u^{p,A}(x,Q^2)$.
\begin{figure}[ht]
\center
\includegraphics[scale=0.3]{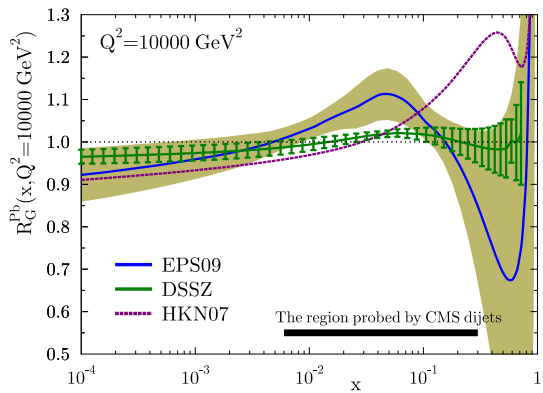}
\caption{The gluon nuclear modification factors from  \texttt{EPS09} (blue line with error band),
\texttt{DSSZ} (green line with error bars) and \texttt{HKN07} (purple dashed line) at
$Q^2 = 10000 \, {\rm GeV}^2$. The approximate range probed by the CMS dijet measurements \cite{CMSprel}
is indicated by the thick black line.}
\label{Fig:EPS09vsDSSZ}
\end{figure}
Throughout this work, we use the \texttt{CT10NLO} \cite{Lai:2010vv} free proton PDFs as a baseline
and employ
the nuclear modifications from the NLO analyses \texttt{EPS09} \cite{Eskola:2009uj}, \texttt{DSSZ} \cite{deFlorian:2011fp},
and \texttt{HKN07} \cite{Hirai:2007sx}.\footnote{Strictly speaking, using \texttt{CT10NLO} PDFs as a baseline for all these
parametrizations 
is not completely consistent as different baseline sets were employed in the original fits. However, this
issue has no consequences with respect to our results here.}
Similar results from the nCTEQ collaboration have also been published \cite{Schienbein:2009kk},
but their final pre-LHC parametrizations are not yet available \cite{Kovarik:2013sya}.

The differences between the nuclear modifications from \texttt{EPS09}, \texttt{DSSZ} and \texttt{HKN07} are sizable
in the case of gluons which play the dominant role in the jet production here and, as we will show,
cause interesting differences in the calculated dijet cross sections. In order to tie these differences to the
differences in the gluon modifications we show in Fig.~\ref{Fig:EPS09vsDSSZ} the gluonic modification, $R_{G}^{\rm Pb}$,
as predicted by different parametrizations at a large scale $Q^2 = 10000 \, {\rm GeV}^2$ relevant for jet production.
Towards small $x$, all three parametrizations tend to agree within the uncertainties but
above $x\backsimeq 10^{-3}$ clear disagreements exist. 
The principal reason for these differences is the exclusion or different implementation of the inclusive
pion production data in deuterium+gold collisions measured at RHIC-BNL \cite{Abelev:2009hx,Adler:2006wg}: While \texttt{EPS09} and \texttt{DSSZ} 
include these data, \texttt{HKN07} omits this type of data. In addition, while in \texttt{EPS09} it was
assumed that the parton-to-pion fragmentation is unaffected by the nuclear environment, \texttt{DSSZ}
included nuclear modifications also in the fragmentation functions \cite{Sassot:2009sh}. Both \texttt{EPS09} and \texttt{DSSZ} 
provide a good description of the RHIC data despite their different fragmentation function philosophy. 
As the dijet cross section is dominated by the low values of dijet mass $M_{\rm dijet}$, 
we can roughly approximate the probed momentum fraction of the nucleus by
$x_1 \approx ( 2p_T^{\rm min, leading}/ \sqrt{s} ) e^{\eta_{\rm dijet}-\eta_{\rm shift}}$ in a conceivable
interval $-2.5 < \eta_{\rm dijet} < 1.5$, obtaining $0.006 \lesssim x_1 \lesssim 0.3$ (note that $\eta_{\rm dijet}$ here
is in the laboratory frame). That is, the
dijet cross sections will probe exactly the controversial range in $R_G^{\rm Pb}(x,Q^2)$ and could therefore
discriminate between the different parametrizations.

\subsection{The smallness of the isospin correction}

In general, the proton+nucleus cross sections are different from 
the proton+proton ones even without the nuclear modifications in PDFs, corresponding to $R_i^A(x,Q^2)=1$, in Eq.~(\ref{eq:nmod}).
This is because of the the isospin effect, i.e. the different relative amount of up and down quarks due
to the presence of neutrons in the nucleus. However,
in the case of jets such effects are, in practice, negligible. This 
follows from the fact that jet production is dominated by the quark+gluon and 
gluon+gluon partonic subprocesses. Writing the convolution between the PDFs and
the partonic jet cross sections, $\hat\sigma_{ij \rightarrow {\rm jet}}$, schematically 
as $ f_i^A \otimes  \hat\sigma_{ij \rightarrow {\rm jet}} \otimes f_j^p$, the contribution
of e.g. the quark-gluon channel can be expressed as
\begin{eqnarray}
& & \sum_i q_i^A  \otimes \hat\sigma_{q_ig \rightarrow {\rm jet}} \otimes g^p  =  
\left[ u^A + d^A + s^A + c^A + b^A \right] \otimes \hat\sigma_{qg \rightarrow {\rm jet}} \otimes g^p  \\
& = & \left[ \frac{Z}{A} \left(u^{p} + d^p + s^p + c^p + b^p \right) + 
             \frac{N}{A} \left(u^{n} + d^n + s^n + c^n + b^n \right) \right] 
\otimes \hat\sigma_{qg \rightarrow {\rm jet}} \otimes g^p \nonumber \\
& = & \left[ \frac{Z}{A} \left(u^{p} + d^p + s^p + c^p + b^p \right) + 
             \frac{N}{A} \left(d^{p} \hspace{0.04cm} + u^p \hspace{0.04cm} + s^p \hspace{0.04cm} + c^p \hspace{0.04cm} + b^p \right) \right] 
\otimes \hat\sigma_{qg \rightarrow {\rm jet}} \otimes g^p \nonumber \\
& = & \sum_i q_i^p \otimes \hat\sigma_{q_ig \rightarrow {\rm jet}} \otimes g^p , \nonumber
\end{eqnarray}
where we used the fact that $\hat\sigma_{q_ig \rightarrow {\rm jet}}$ is independent of the quark flavor
in the absence of electroweak corrections \cite{Dittmaier:2012kx}. In $qq$-subprocesses this is not true but
these processes are not the principal contribution in the jet cross sections. As a consequence, without
nuclear effects in PDFs we have $\sigma^{\rm p+Pb}_{\rm jet} \approx \sigma^{\rm Pb+p}_{\rm jet} \approx \sigma^{\rm p+p}_{\rm jet}$
to a very good approximation in the present framework. A similar disappearance of the isospin effect
in proton+lead collisions was found also in the case of dilepton production \cite{Paukkunen:2010qg}
at the Z boson pole. The lack of an isospin effect is in contrast to e.g. direct photon production where e.g.
$\hat\sigma_{ug \rightarrow \gamma} \neq \hat\sigma_{dg \rightarrow \gamma}$ as the partonic cross sections
are proportional to the square of the electric charge of the quark $e_q^2$ \cite{Arleo:2007js,Arleo:2011gc,Helenius:2013bya}.

\section{NLO Corrections, Scale Dependence and Baseline PDF Errors}

\begin{figure}[ht]
\center
\includegraphics[scale=0.4]{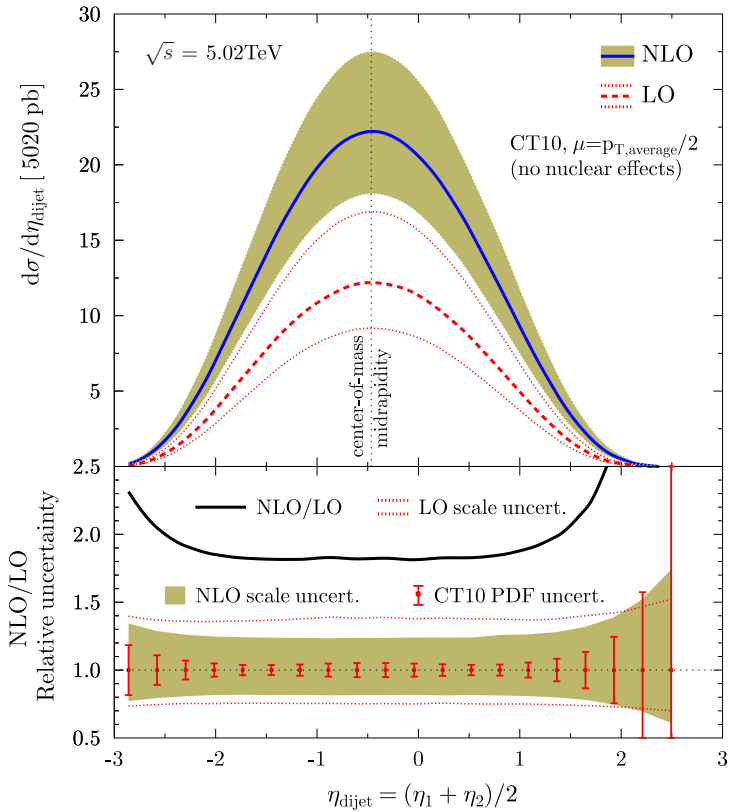}
\caption{Upper panel: The absolute dijet spectrum at LO (dashed red line) and NLO (continuous blue line).
The scale uncertainties are marked by the area enclosed by the dotted lines (LO), and the
shaded band (NLO). The variable $\eta_{\rm dijet}$ is in the laboratory frame, and the vertical dotted line marks the location of the center-of-mass midrapidity. 
Lower panel: The ratio between the NLO and LO calculations (black line). Also shown are the relative  \texttt{CT10} PDF uncertainties (error bars) and the relative scale uncertainties in NLO (shaded band) and LO (band between dotted lines). }
\label{Fig:Spectra}
\end{figure}

\begin{figure}[ht]
\center
\includegraphics[scale=0.4]{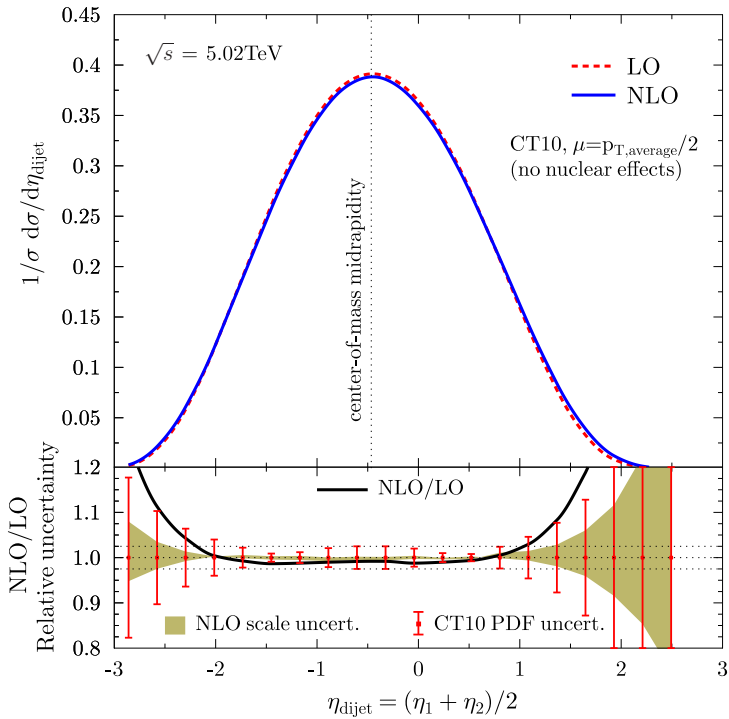}
\caption{Upper panel: The normalized dijet spectrum at LO (dashed red line) and NLO (continuous blue line). 
Lower panel: The ratio between the NLO and LO calculations (black line).
Also shown are the relative  \texttt{CT10} PDF uncertainties (error bars) and the relative scale uncertainties in NLO (shaded band). The dotted lines are to guide the eye for a 2.5\% relative uncertainty band.}
\label{Fig:Spectra22}
\end{figure}

We begin the discussion from the absolute dijet spectrum, shown in Fig.~\ref{Fig:Spectra},
where we present the LO and NLO calculation without nuclear modifications in the PDF, that is,
we take $R_i^{A}(x,Q^2) = 1$ in Eq.~(\ref{eq:nmod}). The scale dependence is estimated by varying
the common factorization and renormalization scale $\mu$ by factors $0.5$ and $2$ around
$\mu = p_T^{\rm average}/2 = (p_T^{\rm leading} + p_T^{\rm subleading})/4$.
This choice of scale is preferred since in the CT10 analysis \cite{Lai:2010vv} the inclusive jet cross
sections were computed by fixing the scales to $p_T/2$. As seen in the Fig.~\ref{Fig:Spectra}, 
the NLO correction is always quite large --- almost a factor of
two at least --- growing strongly towards large $\eta_{\rm dijet}$.
Part of this growth is due to the non-equal transverse momentum cuts: In LO the jets are
back-to-back in the transverse plane, carrying an equal transverse momentum 
$p_{T}^{\rm leading} = p_{T}^{\rm subleading} > p_T^{\rm min, leading} = 120 \, {\rm GeV}$.
The corresponding maximum dijet rapidity is then given by 
$\eta^{\rm max, LO}_{\rm dijet} = \log \left( \frac{\sqrt{s}}{2p_T^{\rm min}} \right) +\eta_{\rm shift} \approx 2.5$.
However, due to the lower $p_T$ cut for the subleading jet, the NLO spectrum
extends above this limit leading to an up-shooting, and, finally to an infinite NLO-to-LO ratio.
Apart from these effects close to the edge of the phase space, the NLO correction turns
out rather flat in the bulk part $-2 \lesssim \eta_{\rm dijet} \lesssim 1$ of the dijet spectrum.
The scale dependence is rather strong, around $30\%$, being generally somewhat larger at LO.
The largeness of the NLO scale uncertainty originates from the low values of dijet mass $M_{\rm dijet} \lesssim 300 \, {\rm GeV}$
which dominate the cross section and where the scale uncertainty is particularly large. Imposing an
additional cut e.g. $M_{\rm dijet} \geq 300 \, {\rm GeV}$ would make the scale uncertainty significantly
smaller.
Another source of uncertainty we consider here stems from the experimental uncertainties of the data
by which the free nucleon PDFs were constrained. The range of such variations are encoded in the PDF error
sets $\{S^\pm_k\}$ of \texttt{CT10}, which we use to calculate the uncertainty $\delta \sigma$ for each cross section
$\sigma$ by
\footnote{Technically, we evaluate the PDF errors by computing the LO part of the cross sections
weighted by multiplicative NLO/LO K-factors computed separately for each rapidity bin using the central set.}
\begin{equation}
\left( \delta \sigma \right)^2 = \frac{1}{4} \sum_k \left[ \sigma \left( S^+_k \right) - \sigma \left( S^-_k \right) \right]^2.
\end{equation}
In comparison to the scale variation, this uncertainty turns out usually much smaller and only at very large $|\eta_{\rm dijet}|$
it becomes comparable with the scale uncertainty, as can be clearly seen from the lower panel of Fig.~\ref{Fig:Spectra}.

A direct experimental measurement of this absolute cross section suffers, however, from the lack of
precise knowledge of the proton+nucleon luminosity in proton+nucleus collisions.
Therefore, in order to avoid resorting to a model-dependent normalization \cite{Miller:2007ri},
it is rather the shape of the theoretical spectra that can be compared rigorously with the measurements.
Since the NLO correction within $-2 \lesssim \eta_{\rm dijet} \lesssim 1$ in Fig.~\ref{Fig:Spectra} 
is, to a good approximation, merely a multiplicative overall factor, the shape of the LO distribution
is already a very good estimate for the full NLO one.
This is demonstrated in Fig.~\ref{Fig:Spectra22}, where the cross sections of
Fig.~\ref{Fig:Spectra} have been normalized by the corresponding cross sections integrated over $\eta_{\rm dijet}$.
As the size of the NLO corrections within $-2 \lesssim \eta_{\rm dijet} \lesssim 1$ are already very small, we would expect 
that the NNLO corrections will be negligible as they mainly affect the
overall normalization rather than the shape of the rapidity spectra \cite{Ridder:2013mf}. This expectation
is also supported by the dramatically reduced sensitivity to the scale variations in comparison to the
absolute cross sections. Indeed, contrary to the absolute spectrum in Fig.~\ref{Fig:Spectra},
the PDF uncertainties typically dominate in the normalized spectrum.

\begin{figure}[ht]
\center
\includegraphics[scale=0.4]{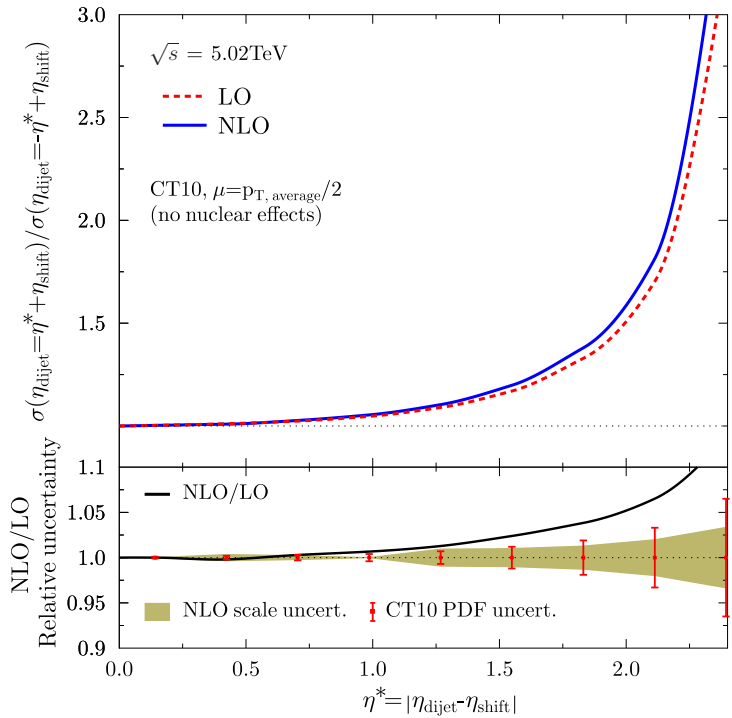}
\caption{Upper panel: The forward-to-backward ratio at LO (dashed red line) and NLO (continuous blue line)
as a function of the variable $\eta^*=|\eta_{\rm dijet}-\eta_{\rm shift} |$.
Lower panel:
The ratio between the NLO and LO calculations (black line).
Also shown are the relative  \texttt{CT10} PDF uncertainties (error bars) and the relative scale uncertainties in NLO (shaded band). }
\label{Fig:Spectra33}
\end{figure}

Another observable that avoids measuring the absolute normalization is the
ratio between the yields in the forward and backward rapidity bins around the center-of-mass
midrapidity, as shown in Fig.~\ref{Fig:Spectra33}. The effect of the asymmetric rapidity acceptance is
particularly visible in this ratio: Due to the wider rapidity acceptance in the forward direction
the cross sections are there larger in comparison to the backward direction, and consequently
the forward-to-backward ratio is always above unity. Indeed, had the acceptance been symmetric
around the center-of-mass midrapidity $\eta=\eta_{\rm shift}$, the spectrum would have been almost symmetric due to the
lack of isospin effects (as discussed earlier and as we have verified numerically), and, consequently,
the ratio almost exactly unity both
in LO and NLO irrespectively of the scale choices. In spite of the presence of this asymmetry, the
perturbative convergence seems to be still well under control as the scale dependence is less than 5\%
in the  experimentally achievable range in $|\eta_{\rm dijet}|$. In general, both the scale dependence
and the PDF uncertainties are reduced from those in the normalized spectrum.

\begin{figure}[h]
\center
\includegraphics[scale=0.4]{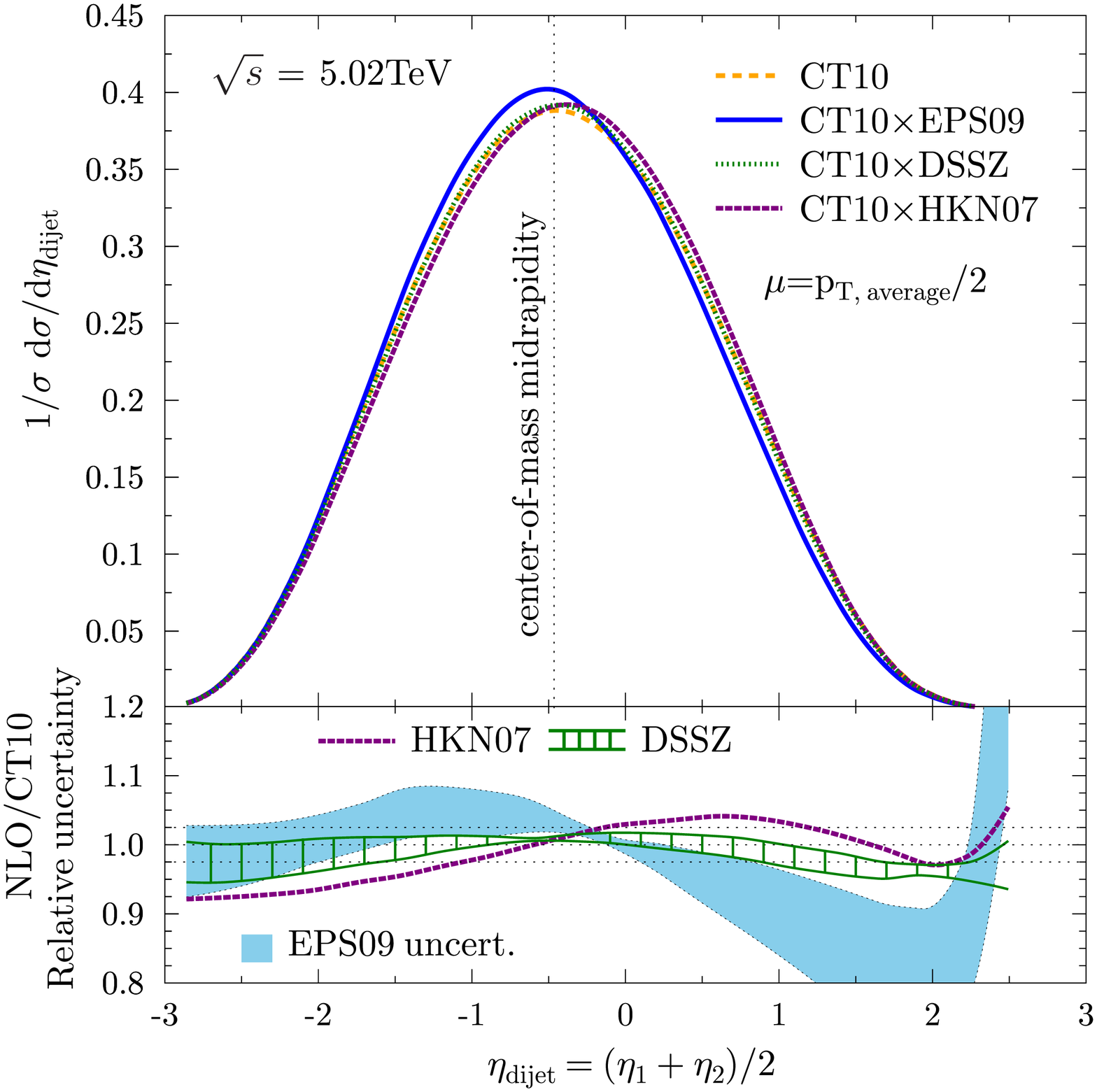}
\caption{Upper panel: The normalized NLO dijet spectrum without nuclear effects in PDFs (orange dotted line) and
with nuclear modifications from \texttt{EPS09} (blue solid line), \texttt{DSSZ} (green dotted line),
and \texttt{HKN07} (purple dashed line). 
Lower panel: The normalized NLO dijet spectrum with nuclear effects divided by the corresponding calculation without the nuclear effects.  The blue band corresponds to the
\texttt{EPS09} uncertainty, and the green hatched band is the
\texttt{DSSZ} uncertainty range. The dotted lines mark again the 2.5\% baseline uncertainty as in Fig.~\ref{Fig:Spectra22}.
}
\label{Fig:neffects}
\end{figure}

As the dijet cross sections are steeply falling functions of ${M_{\rm dijet}}$, all
the observables presented here are dominantly sensitive to the low $M_{\rm dijet} \sim 240 \, {\rm GeV}$,
where the electroweak corrections \cite{Dittmaier:2012kx} are not important. Also, the
rapidity dependence of the non-perturbative corrections, as given by the Monte-Carlo event
generators, appears rather mild \cite{Aad:2011fc} and should therefore mostly cancel 
out in the normalized spectrum and even more completely  in the forward-to-backward ratio. 
All in all, the shape of the dijet spectrum and the forward-to-backward ratio appear as promising,
precision-observables to study.

\section{Nuclear Modifications in the Dijet Spectrum}

\begin{figure}[ht]
\centering
\includegraphics[scale=0.305]{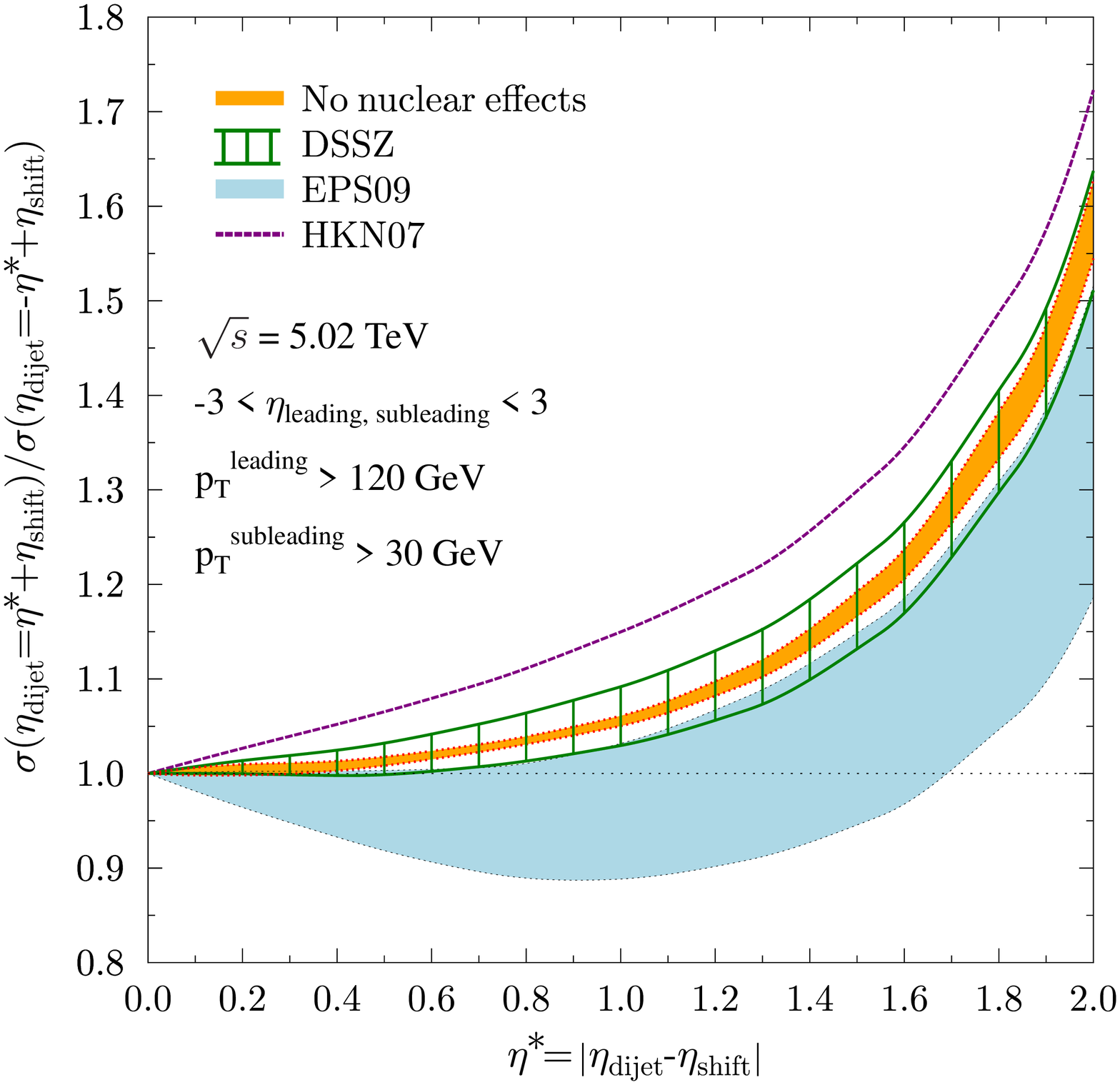}
\hspace{-0.6cm}
\includegraphics[scale=0.305]{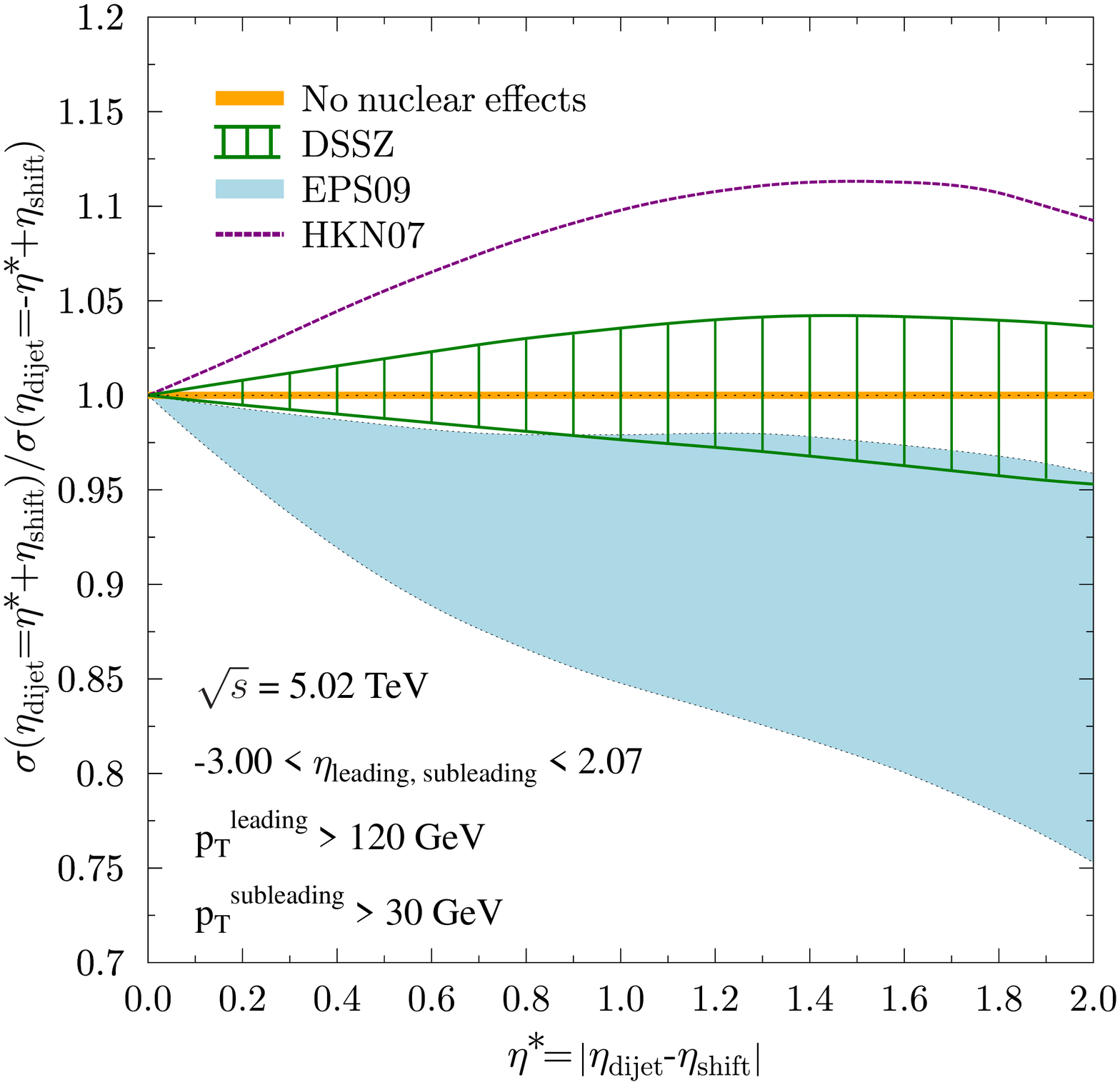}
\caption{Left: The forward-to-backward ratio without nuclear effects in PDFs (thin orange band for the \texttt{CT10} PDF uncertainties)
and with nuclear modifications from \texttt{HKN07} (purple dashed line), \texttt{EPS09} (light blue band)
and \texttt{DSSZ} (green hatched band), using the 
pseudorapidity acceptance $|\eta^{\rm leading, subleading}| < 3$. %and $p_T^{\rm min, leading} = 120 \, {\rm GeV}$. 
Right: The same but for the acceptance $-3 < \eta^{\rm leading, subleading} < 2.07$.}
\label{Fig:neffects2}
\end{figure}

Having now discussed the reliability of the NLO calculations and the uncertainties of the baseline
against which the experimental data and nuclear modifications can be compared to, we turn to the effects induced by the
nuclear modifications in the PDFs, $R_i^A(x,Q^2)\neq 1$. As argued, the $\eta_{\rm dijet}$
dependence of the dijet spectrum should reflect the $x$ dependence of the $R_{G}^A(x,Q^2)$
PDF plotted in Fig.~\ref{Fig:EPS09vsDSSZ}. The results of the NLO calculations with \texttt{CT10NLO}
PDFs, modified by the nuclear effects from \texttt{EPS09}, \texttt{HKN07} and \texttt{DSSZ}, are
presented in Fig.~\ref{Fig:neffects} in the case of the normalized spectrum. 
The shapes of the distributions evidently become distorted from those expected
without nuclear modified PDFs. The mutual ordering of the \texttt{EPS09},
\texttt{HKN07} and \texttt{DSSZ} results seen here roughly follows the ratios $R_{G}^A(x,Q^2)$ in Fig.~\ref{Fig:EPS09vsDSSZ},
although the effects are smoothed out by the integration over the transverse momenta of
the jets, and by the valence-quark contributions which gradually take over towards forward $\eta_{\rm dijet}$.
In any case, the mutual differences between the predictions are larger than the baseline 2.5 \% uncertainty at 
$-2 \lesssim \eta_{\rm dijet} \lesssim 1$.
The variations between the \texttt{EPS09}, \texttt{HKN07} and \texttt{DSSZ} results become even more pronounced
in the forward-to-backward ratio shown in the left-hand panel of Fig.~\ref{Fig:neffects2}. In the case of \texttt{EPS09}
the depletion in the forward rapidity (the EMC effect) and the enhancement in the backward direction (antishadowing)
enhance the total effect, causing a drastic difference in comparison to the calculation 
with \texttt{DSSZ} or \texttt{HKN07} or without nuclear modifications.
Thus, these observables are particularly suitable for testing the nuclear gluon PDFs.

Although the baseline uncertainty in the forward-to-backward ratio is already small, and dominated by the 
free proton PDF errors, it can be made negligible by changing the pseudorapidity acceptance of the
individual jets such that it is symmetric in the proton-nucleon center-of-mass frame. Setting the cut 
$-3 < \eta^{\rm leading, subleading} < 2.07$ in the laboratory frame corresponds to accepting jets, in LO,
within $|\eta^*_{\rm leading, subleading}| < 2.535$ in the proton-nucleon center-of-mass frame. The effect
of imposing this cut is shown in the right-hand panel of Fig.~\ref{Fig:neffects2}. In the absence of isospin effects
the baseline calculation is simply unity while the nuclear effects in the PDFs clearly stand out.
\begin{figure}[ht]
\centering
\includegraphics[scale=0.305]{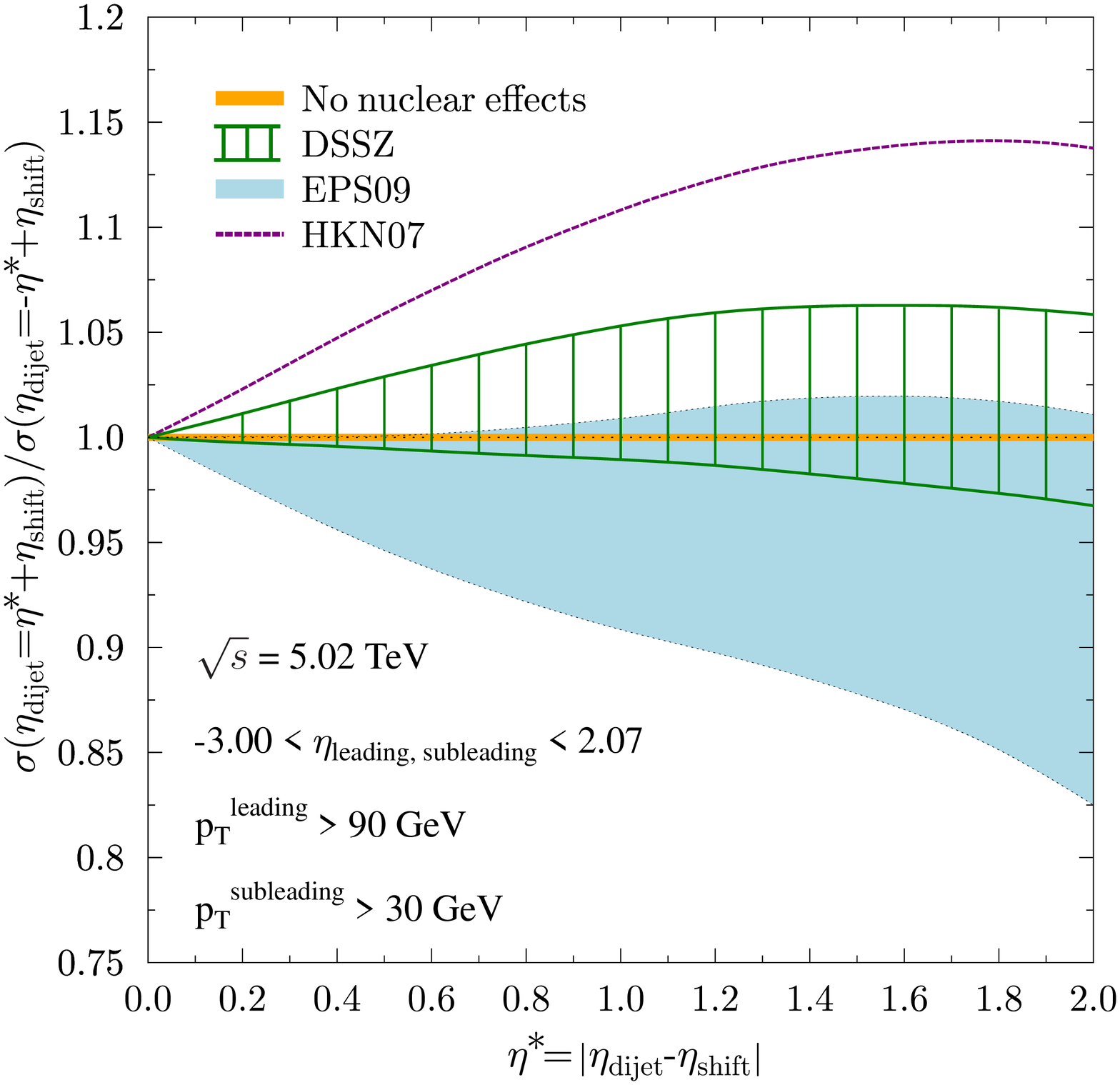}
\hspace{-0.6cm}
\includegraphics[scale=0.305]{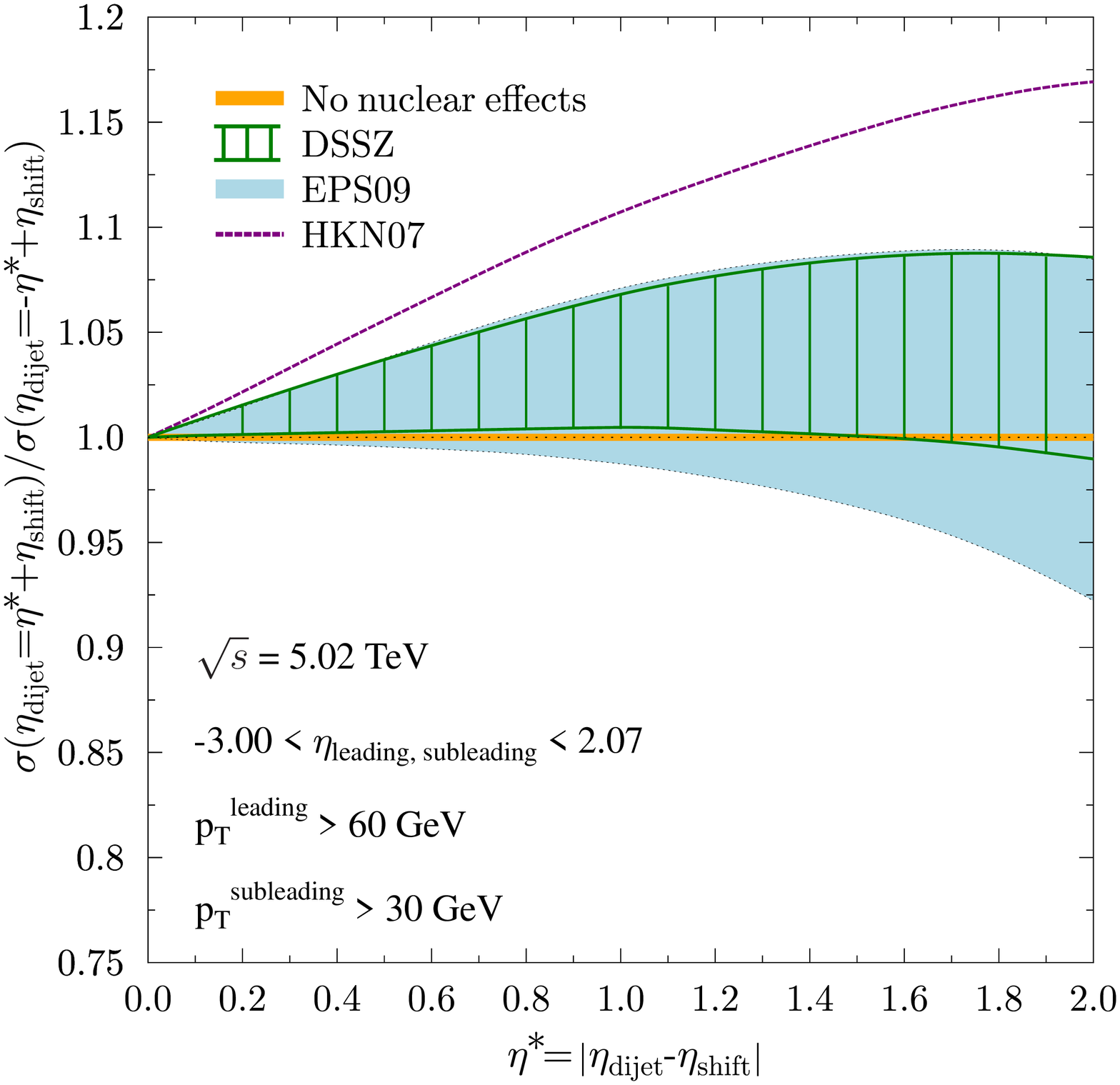}
\caption{As the right-hand panel of Fig.~\ref{Fig:neffects2}, but with
lower cuts for $p_T^{\rm leading}$.}
\label{Fig:neffects3}
\end{figure}
The qualitative behaviour can be understood by
\begin{equation}
 \frac{\sigma (\eta_{\rm dijet}=\eta_{\rm dijet}^* +\eta_{\rm shift} )}{\sigma (\eta_{\rm dijet}=-\eta_{\rm dijet}^* +\eta_{\rm shift} )} \approx
\frac{R_{\rm gluon}^A \left(\xi_1, Q^2  \right)}
     {R_{\rm gluon}^A \left(\xi_2, Q^2  \right)},
\end{equation}	
where $\xi_1 = {(2p_T^{\rm min, leading}}/{\sqrt{s}})e^{\eta_{\rm dijet}^*}$,
      $\xi_2 = {(2p_T^{\rm min, leading}}/{\sqrt{s}})e^{-\eta_{\rm dijet}^*}$.
With $p_T^{\rm min, leading} = 120 \, {\rm GeV}$,
the center-of-mass midrapidity $\eta^*_{\rm dijet}=0$ sits at $\xi_1(\eta^*_{\rm dijet}=0) = \xi_2(\eta^*_{\rm dijet}=0) \approx 0.05$,
which coincides with the antishadowing peak of the \texttt{EPS09} in Fig.~\ref{Fig:EPS09vsDSSZ}.
By increasing $\eta_{\rm dijet}^*$, $\xi_1$ grows, $\xi_2$ gets smaller, and
consequently the ratio is taken between the EMC depletion and the antishadowing
enhancement in $R_{G}(x,Q^2)$, which roughly explains the suppression seen in
the right-hand-panel of Fig.~\ref{Fig:neffects2}.
Although we have here considered only very specific kinematical conditions, it is
rather easy to understand the systematics of e.g. changing the center-of-mass energy $\sqrt{s}$ or
varying the leading jet transverse-momentum cut $p_T^{\rm min, leading}$. To this
end, Fig.~\ref{Fig:neffects3} shows two examples of the forward-to-backward
ratio with lower $p_T^{\rm min, leading}$.
The largest sensitivity to such systematics is 
evidently in the \texttt{EPS09} results for which the ratio tends to grow when the
$p_T^{\rm min, leading}$ is decreased. However, from the relation above this effect
can be easily understood: For instance, with $p_T^{\rm min, leading} = 60 \, {\rm GeV}$
we have $\xi_1(\eta^*_{\rm dijet}=0) = \xi_2(\eta^*_{\rm dijet}=0) \approx 0.02$,
and the forward-to-backward ratio begins to reflect the division of the antishadowing enhancement
by the small-$x$ shadowing in $R_{G}(x,Q^2)$ making the ratio eventually larger than unity.

\section{Summary}

In conclusion, we have performed a detailed study of the perturbative QCD expectations for dijet production in proton+lead collisions
with a special focus on the measurements recently performed by the CMS collaboration \cite{CMSprel}. By studying the size
of the NLO corrections, sensitivity to the scale variations and free proton PDF errors, we have found that
such baseline uncertainties remain below few percents in the bulk part $-2 < \eta_{\rm dijet} < 1$ of the pseudorapidity
spectrum which is normalized by the corresponding total dijet cross section. By taking ratios of the dijet yields between
the positive and negative sides of the midrapidity, the baseline uncertainties become even more suppressed, and, can
be made practically negligible by suitably symmetrizing the rapidity acceptance. Expectations derived from the latest
NLO parametrizations of the nuclear PDF modifications contain notable mutual differences which can be tracked down to
the qualitatively very different modifications of the gluon PDF at large $x$. The forthcoming data from the CMS collaboration
are expected to be accurate enough \cite{CMSprel} to make conclusions regarding the validity of the factorization for the
jet production, the assumption of the isospin symmetry, and hopefully to provide completely new, stringent, constraints for
the nuclear modifications of the gluon PDFs.

\section*{Acknowledgments}

We thank Yen-Jie Lee from the CMS Collaboration, Francois Arleo, N\'estor Armesto and Guilherme Milhano for useful discussions.
H.P. and K.J.E. acknowledge the support from the Academy of Finland, Project No. 133005.
C.A.S. is supported by European Research Council grant HotLHC ERC-2011-StG-279579, by Ministerio de Ciencia e 
Innovacion of Spain under grant No. FPA2009-06867-E, and by Xunta de Galicia.

\end{document}